\documentstyle[aps,preprint,epsf,epsfig,rotate,prl]{revtex}
\newcommand{\be}{\begin{equation}}
\newcommand{\ee}{\end{equation}}
\newcommand{\ba}{\begin{eqnarray}}
\newcommand{\ea}{\end{eqnarray}}

\draft        
\begin{document}             
\title{Supercurrent and noise in point contact 
between two different superconductors}
\author{S. K. Yip}
\address{Institute of Physics, Academia Sinica, Nankang, Taipei 11529, Taiwan}
\maketitle

\begin{abstract}

We show that, for a quantum point contact between two superconductors
of different gap magnitudes, Andreev bound states do not exist
for certain phase differences and gap ratios.  
Continuum states may dominate the transport, and the supercurrent
noise is qualitatively different from the case of equal gaps.


\end{abstract}

With nanotechnology an emerging subject, there has
been much effort in understanding electrical transport in
quantum point contacts \cite{meso}.
A particular interesting sub-field is the study of contacts between
two superconductors.  Their transport
properties such as current-phase ($I-\phi$) relationships and
current-voltage dependences have been studied theoretically
\cite{Averin95,AverinB96,Cuevas99}
as well as measured experimentally
\cite{Koops96,Goffman00}.
There are further theoretical works on the important issue
of current fluctuations (noise). 
\cite{Cuevas99,AverinL96,Martin96,Naveh99}
All these theoretical papers are for the case where
the two superconducting gaps $|\Delta|$ are equal.
In the physical pictures developed in these papers,
a particularly important role is played by the Andreev bound
states that form between the two superconductors.
For a short junction with a single conduction channel of 
transmission probability $D$,
there is a pair of bound states at $\pm \epsilon_b$ where
$\epsilon_b = |\Delta| \left[ 1 - D {\rm sin}^2 (\phi/2) \right]^{1/2} $.
\cite{Furusaki90,Beenakker91}
Here $\phi$ is the phase difference between the two superconductors.
In equilibrium, the net supercurrent $I$ is carried entirely
by this bound state, and at zero temperature,
$I$ is given simply by $ I = - { 2e \over \hbar} {\partial \epsilon_b
\over \partial \phi}$, while at a finite temperature $T$,
$I$ is reduced by a factor ${\rm tanh} \left( {\epsilon_b \over 2 T} \right)$
arising from the occupation numbers at $ \pm \epsilon_b$.
The response of these bound states also play an important 
role in determining the non-equilibrium electron
transport such as $IV$ characteristics and the
junction behavior under microwave irradiation.
\cite{Averin95,AverinB96,Cuevas99}

These Andreev bound states also play a major role in determining
the supercurrent noise of the point contact.
\cite{Cuevas99,AverinL96,Martin96,Naveh99}
An important source of the 
equilibrium supercurrent noise 
arises from the fact that these bound states can either
be occupied or empty due to thermal fluctuations.\cite{AverinL96,Martin96}
As a result, there are two strong peaks of the current
noise at zero frequency and $\omega = 2 \epsilon_b$.

In this paper we study the bound state, current-phase relationship,
as well as noise of a quantum contact between two superconductors
with unequal gaps.  We shall show that, in some regions of parameter
space, the situation is dramatically different from the case
where the gaps are of equal magnitude.  For certain gap ratios
and phase difference, the Andreev bound states do not exist at all.
Instead of having bound state carrying all the current, the 
{\it continuum} states
with energy between the two gaps can carry a substantial, sometimes
even the dominating, part of the supercurrent.  
Interestingly the total supercurrent, and hence the current-phase
relationship, is not significantly
changed (except the overall magnitude) even in the case of unequal gaps.
However,
when the bound states do not exist,
 the supercurrent {\it noise} is qualitatively different.
  In particular, there is a substantial reduction in noise
level for all frequencies below twice the smaller gap.  The 
zero frequency and $2 \epsilon_b$ peaks are entirely absent.
It should be possible to perform measurements such as STM with 
a Pb tip on Al to verify
the predictions in the present work.
  
We shall then consider two s-wave superconductors in contact through
a barrier with transmission coefficent $D$.  Without loss in generality
we shall take the left (right) superconductor with smaller (larger)
gap magnitude $|\Delta_l|$ ( $|\Delta_r|$),
and take the phase difference $\phi$ between $0$ and $\pi$.
 One can easily
adapt the quasiclassical formalism \cite{Zaitsev84}
to the present (one conduction
channel) case, and obtain the Green's functions
$\hat g_{l,r}(p, \epsilon)$ and $\hat g_{l,r} ({\underline p}, \epsilon) $
for positive ($p$) and negative (${\underline p} = - p$) momenta
and at a point near and to the left and right (subscript $l$ and $r$) of the
barrier.  It is convenient to express the answers 
\cite{Zaitsev84,Yip97} in terms of 
$\hat s_{l,r} \equiv \hat g_{l,r} (p) + \hat g_{l,r} (\underline p)$
and 
$\hat d_{l,r} \equiv \hat g_{l,r} (p) - \hat g_{l,r} (\underline p)$.
One obtains $\hat d_l = \hat d_r = \hat d$ with
\be
\hat d = { i D \over 2 \pi} \left[ \hat g_{r,\infty} , 
   \hat g_{l,\infty} \right] \ / \ {\cal D}
\label{d}
\ee
and
\be
\hat s_l = \left( ( 2 - D) \hat g_{l,\infty} +  
   D \hat g_{r,\infty} \right) \ / \ {\cal D}
\label{s}
\ee
where
$ {\cal D} (\epsilon) \equiv 1 + { D \over 4 \pi^2}
\left( \hat g_{l,\infty} - \hat g_{r,\infty} \right)^2 $.
 Here $\hat g_{l,r,\infty} = 
 - \pi { \epsilon \hat \tau_3 - \hat \Delta_{l,r}
\over \sqrt{ |\Delta_{l,r}|^2 - \epsilon^2}  }$
 are the quasiclassical
Green's function at positions far away from the interface
into the left (right) superconductor.
The above equations hold for both the retarded ($R$) and 
advanced ($A$) Green's functions with
$\epsilon \to \epsilon \pm i 0_{+}$ and we have left out
their superscripts temporarily for simplicity.  Here 
$0_+$ is an infinitesimal positive number.
Explicitly, we have
\be
\hat d = i \pi D
{\epsilon \hat \tau_3 (\hat \Delta_r - \hat \Delta_l) -
i |\Delta_r| |\Delta_l| {\rm sin} (\phi) \hat \tau_3
\over 
Q (\epsilon) }
\label{dex}
\ee
where $Q(\epsilon) \equiv 
\sqrt{|\Delta_{l}|^2 - \epsilon^2} 
\sqrt{|\Delta_{r}|^2 - \epsilon^2} \ 
{\cal D} (\epsilon)$, {\it i.e.}
\be
Q (\epsilon) =  ( 1 - {D \over 2} )
\sqrt{|\Delta_{l}|^2 - \epsilon^2} 
\sqrt{|\Delta_{r}|^2 - \epsilon^2}
\ - \ 
{ D \over 2}
\left( \epsilon^2 - |\Delta_r| |\Delta_l| {\rm cos} (\phi)  \right)
\label{Q}
\ee

 Andreev bound state exists when 
\be
Q(\epsilon) = 0
\label{bs}
\ee
 for some $|\epsilon| < |\Delta_l| \le |\Delta_r|$. 
The solutions to
eq (\ref{bs}) always come in $\pm \epsilon_b$ pairs and we
shall focus on $\epsilon_b > 0$.
Eq (\ref{bs}) can be rewritten as a quadratic equation in $\epsilon^2$ 
(by putting one of the terms of (\ref{Q}) to the other
side and taking the square).
  It can be shown that one of the roots
is larger than $|\Delta_l|^2$ and thus must be rejected.
Thus for a given phase difference, there
is at most one pair ($\pm \epsilon_b$) of bound states.
The remaining root must be substituted back into eq (\ref{bs})
to check that it is indeed a solution.
   We found that as 
 the ratio in gap magnitude increases, there is 
an increasing region near $\phi = 0$ where the bound states
cease to exist.  
Examples are shown in Fig. \ref{fig:d0.9} and {\ref{fig:z0.5}.
 The value of $\phi$ where the bound
states first appear is given by $\phi_c = {\rm cos}^{-1}
(|\Delta_l| / |\Delta_r|) $ 
( $\epsilon = |\Delta_l| $ is a solution to
eq (\ref{bs}) at ${\rm cos} \phi = |\Delta_l| / |\Delta_r| $).
This critical phase difference $\phi_c$ is independent of $D$,
and $\phi_c \to \pi/2$ when $|\Delta_l| << |\Delta_r|$.

To understand this $\phi_c$, it is convenient 
\cite{Beenakker91} to imagine two
small normal regions between the two superconductors surrounding
the barrier (thus a 
${\rm S}_l {\rm N}_l {\rm I} {\rm N}_r {\rm S}_r$ junction.  Here
${\rm S}_{l,r}$ are the two superconductors, ${\rm I}$ the barrier
and ${\rm N}_{l,r}$ are the two normal regions.). Quasiparticles
and holes in ${\rm N}_{l,r}$ are transmitted and reflected by
the barrier ${\rm I}$ while Andreev reflected if they incident
on the superconductors.  Recall that for a particle (hole)
incident on a superconductor, the Andreev reflected hole (particle)
acquires a phase
 $ - {\rm cos}^{-1} \left( { \epsilon \over |\Delta| } \right)$
in addition to that from the phase of the order parameter
$\mp {\rm arg} (\Delta)$.  Now one can easily understand
$\phi_c$ for the case with perfect transmission.  In this
case one can ignore ${\rm I}$ and identify ${\rm N}_{l,r}$.
Let us take, without loss in generality, $\phi_r = \phi$ and
$\phi_l = 0$.  One can readily see that one has an Andreev
bound state at $\epsilon = |\Delta_l|$ at $\phi = \phi_c$ consisting of 
a hole propagating to the right and a particle propagating to
the left.  In this case there is no phase shift at the Andreev
reflection at the left superconductor (since $\epsilon = |\Delta_l|$),
while the two contributions at the reflection at the right
superconductor cancels 
($ \phi_r - {\rm cos}^{-1} ( { \epsilon \over |\Delta_r| } )
= 0$ ).

Note that for the above case, the phase shift does 
{\it not} cancel for an incident particle (reflected hole)
to the right superconductor.  
Since reflection at the barrier I can in general generate
this amplitude, it may be thus surprising
at first sight that the bound state at  $\epsilon = |\Delta_l|$
still exists when $D \ne 1$.  However, a careful analysis
(along the lines of \cite{Beenakker91}) shows that 
the bound state still exists but with the amplitude of
the right moving particle in ${\rm N}_r$ vanishes identically.
The amplitude of the particle incident from the left of
the barrier (in ${\rm N}_l$) and the amplitude of the particle
incident from the right of it (in ${\rm N}_r$) are in
such a ratio that the outgoing, right moving particle amplitude
in ${\rm N}_r$ still vanishes exactly.

The supercurrent is given by
\be
I = - { e \over 2 \pi^2 i \hbar} \ 
\int \ d \epsilon { 1 \over 4} {\rm Tr}_4 
\left\{ \hat \tau_3 \hat d ^{R-A} \right\} \ f(\epsilon)
\label{cur}
\ee
where we have used the short-hand $\hat d^{R-A} \equiv
\hat d^R - \hat d^A$ and $f(\epsilon)$ is the Fermi function.
Explicitly, we find
\ba
I &=& 
{ e D |\Delta_r| | \Delta_l| {\rm sin} \phi
  \over \hbar \  \epsilon_b}
  { \left[ 1 \over ( 1 - { D \over 2} ) 
  ( { \alpha_r \over \alpha_l } + { \alpha_l \over \alpha_r } )
    + D  \right] }
  {\rm tanh} ( {\epsilon_b \over 2 T} ) \nonumber \\
 & &   + 
{ e D | \Delta_r| |\Delta_l| {\rm sin} \phi \over \pi \hbar}
  \ \int_{|\Delta_l|}^{|\Delta_r|}  d \epsilon
  { ( 1 - { D \over 2} ) \sqrt{ |\Delta_r|^2 - \epsilon^2}
                 \sqrt{ \epsilon^2 - |\Delta_l|^2 } \ 
    {\rm tanh} ( \epsilon / 2 T )
     \over
         ( 1 - { D \over 2} )^2   (|\Delta_r|^2 - \epsilon^2)
                 ( \epsilon^2 - |\Delta_l|^2 ) 
         + 
           ({ D \over 2})^2   [ |\Delta_r| |\Delta_l| {\rm cos} \phi 
                         - \epsilon  ^2 ]^2        }
\label{cur2}
\ea
where we have defined
$\alpha_{l,r} = \sqrt{ |\Delta_{l,r}|^2 - \epsilon_b^2 } $.
The first term, $I_b$,  
due to the bound states, is to be included only when these states exist. 
$I_b$ can also be found from
$ - { 2 e \over \hbar} { \partial \epsilon_b \over \partial \phi}$.
Examples of $I_b$ were plotted in Fig \ref{fig:d0.9} and \ref{fig:z0.5}.
$I_b$, however, is not the only contribution to the current
(as in the case for equal gaps).  
$\hat d^{R-A}$ is also finite for $| \Delta_l| < \epsilon < |\Delta_r|$
(though still vanishes for $|\epsilon| > |\Delta_r|$ ).
This (and thus the second term of eq (\ref{cur2}))
 represents the contribution from continuum states
with energies in the above range incident from the left superconductor and
Andreev reflected from the right one.
These states carry a finite current when $\phi \ne 0$.
For low transmission, the continuum states contribution
can actually dominate.
Two examples of the various contributions to the current
are shown in Fig. \ref{fig:cura}.   It turns out that
the dependence of the total current on $\phi$ is not
much different from the case with equal gaps,
except the magnitude is modified.

We finally turn to the current noise of our junction.
We shall study the quantity  \cite{notes}
\be
S_a (\omega) \ \equiv \ \int \ d t  e^{i \omega t}
\left( < \hat I (t) \hat I (0) > - < \hat I > ^2 \right)
\label{sa}
\ee
$S_a$ represents the spectral density for absorption by the junction
of energy quanta $\omega$.  $S_a$ can be written ({\it c.f.} \cite{Khlus87})
in terms of the envelope functions \cite{MRS} for the Green's function
near the interface (see Ref \cite{Appendix}) as
\be
S_a(\omega) = { e^2 \over \hbar} \int \ {d \epsilon \over 2 \pi}
    \left( 1 - f(\epsilon+ \omega) \right) f (\epsilon) \ 
   K (\epsilon) 
\label{Sa}
\ee
where
\ba
K (\epsilon) &\equiv&
 { 1 \over 2} \{ {\rm Tr}_4  [
          \hat \tau_3 \hat C_{++}^{R-A} ( \epsilon+ \omega ) 
             \hat \tau_3 \hat C_{++}^{R-A} ( \epsilon ) 
         +  \hat \tau_3 \hat C_{- -}^{R-A} ( \epsilon+ \omega ) 
            \hat \tau_3 \hat C_{- -}^{R-A} ( \epsilon )  \nonumber \\
 & &  \qquad -  \hat \tau_3 \hat C_{+-}^{R-A} ( \epsilon+ \omega ) 
      \hat \tau_3 \hat C_{-+}^{R-A} ( \epsilon ) 
   - \hat \tau_3 \hat C_{-+}^{R-A} ( \epsilon+ \omega ) 
      \hat \tau_3 \hat C_{+-}^{R-A} ( \epsilon ) ] \}  
\ea

This formula is valid on either side of the junction, and we shall
choose to do this on the left (and correspondingly leave out
the subscripts $l$ in $\hat C$'s.)
$\hat C_{++}$ and $\hat C_{--}$ can be expressed in terms of
$\hat s$ and $\hat d$ given earlier via
$\hat C_{\pm \pm} = \pm { i \over 2} + { 1 \over 4 \pi} ( \hat s_l \pm \hat d)$.
$\hat C_{+-}$ and $\hat C_{-+}$  are given by 
({\it c.f.} \cite{Khlus87}, see also \cite{Appendix})
$\hat C_{+-} = { i \over 2 } r^{*}
  \left( \hat 1 - { i \over \pi} \hat g_{l,\infty} \right)  / {\cal D}$
and 
$\hat C_{-+} = - { i \over 2 } r
  \left( \hat 1 + { i \over \pi} \hat g_{l,\infty} \right)  / {\cal D}$.
Here $r$ is the reflection coefficient for normal quasiparticles
incident from the left of the barrier, and
$|r|^2 = 1 - D$.  

With these $\hat C$'s, the expression for $S_a$ can be written as
eq (\ref{Sa}) with
$K = K_1 + K_2$ where
\ba
K_1 (\epsilon, \omega) &\equiv&
  - 4 ( 1 - D) \
   {\rm Im} \left[ { 1 \over {\cal D}} \right]_{\epsilon + \omega}
       {\rm Im} \left[ { 1 \over {\cal D} } \right]_{\epsilon}
   \nonumber \\
  & &  + D ^2 \
    \left\{ \ 
  \left( \tilde \mu_{l,\epsilon+ \omega} \tilde \mu_{l,\epsilon}
      + \mu_{l,\epsilon+\omega} \mu_{l,\epsilon}  \  |\Delta_l|^2 \right) +
   \left( l \to  r \right) \right\}   
     \label{K1} \\
 & & + D ( 2 - D) \  
    \left\{
  ( \tilde \mu_{l,\epsilon + \omega}
     \tilde \mu_{r,\epsilon}
   +  \tilde \mu_{r,\epsilon + \omega}
      \tilde \mu_{l,\epsilon}  ) \ 
    +   
     ( \mu_{l,\epsilon + \omega}
       \mu_{r,\epsilon} 
    +    \mu_{r,\epsilon + \omega}
             \mu_{l,\epsilon}  )
    (  |\Delta_l| |\Delta_r | {\rm cos} \phi   ) \ 
 \right\}           \nonumber
\ea
and
\ba
K_2 (\epsilon, \omega) &\equiv&
  D^2 
   \{
 {\rm Im} \left[ {\epsilon \over Q } \right]_{\epsilon + \omega}
 {\rm Im} \left[ {\epsilon \over Q  } \right]_{\epsilon}
    ( |\Delta_r|^2 + |\Delta_l|^2  - 2 |\Delta_r| |\Delta_l| {\rm cos} \phi )
   \nonumber \\
   & & + {\rm Im} \left[ { 1 \over Q} \right]_{\epsilon + \omega}
     {\rm Im} \left[ {1 \over Q  } \right]_{\epsilon}
    ( |\Delta_r|^2 |\Delta_l|^2 {\rm sin}^2 \phi )  
    \}
\label{K2}
\ea
Here we have introduced the short hands 
$\mu_{l,r} (\epsilon) =   {\rm Im} \left[ { 1 \over 
            \sqrt{ |\Delta_{l,r}|^2 - \epsilon^2 } \
          {\cal D} (\epsilon) } \right] $
and
$\tilde \mu_{\l,r} (\epsilon) \equiv {\rm Im} \left[ { \epsilon \over 
            \sqrt{ |\Delta_{l,r}|^2 - \epsilon^2 } \
          {\cal D} (\epsilon) } \right] $.
On the right hand sides of eq (\ref{K1}) and (\ref{K2}),
 the subscripts $\epsilon$ or $\epsilon + \omega$
specify where the functions are to be evaluated and
all $\epsilon$'s are understood to be $\epsilon + i \eta$.  
Here $\eta$ is a damping coefficient. \cite{AverinL96,Martin96}

An example of the current-noise is as shown in Fig \ref{fig:noise}.
At increasing gap ratios, $\epsilon_b$ moves towards the gap edge
$|\Delta_l|$.  Correspondingly the peaks at $\omega = 0$ and
$\omega = 2 \epsilon_b$ decreases.  When the bound states
no longer exist, these two peaks are entirely absent and
the noise becomes low for all frequencies below twice
the smaller gap.  This happens in particular for small phase
differences and large gap ratios.  Note also that the
contribution from continuum states is strongly suppressed
when $ T << |\Delta_l|$.

In conclusion, we have studied the Andreev bound states and
supercurrent for a quantum point contact between two superconductors
with different gap magnitudes.  We showed in particular that
the current noise can be qualitatively different from the 
case where the superconductors are identical.

This research was supported by the National Science Council
of Taiwan grant number NSC90-2112-M-001-061 and 
 NSC91-2112-M-001-063.

\begin{figure}[h]
\epsfig{figure= 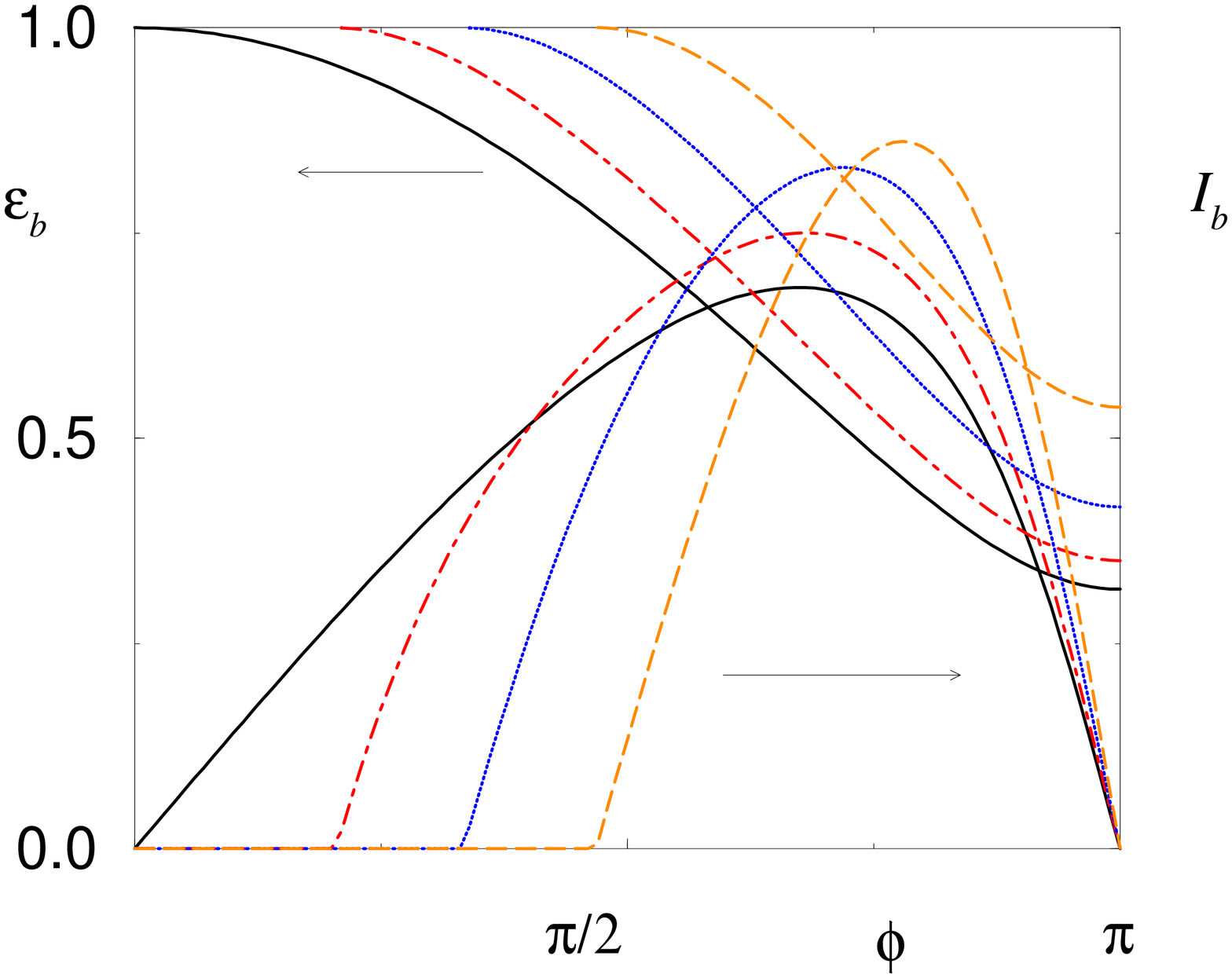,width=5in,angle=-90}
\vskip 0.4 cm
\begin{minipage}{\textwidth}
\caption[]{ Bound state energy $\epsilon_b$ (in unit
of $|\Delta_l|$, left axis), and 
the bound state contribution to the current
(in unit of $e |\Delta_l| / \hbar $, right axis) 
as a function of $\phi$.  $D = 0.9$,
while $|\Delta_r|/|\Delta_l|$'s are given by
$1.0$ (full line), $1.25$ (dot-dashed),
$2.0$ (dotted) and $10$ (long-dashed). }

\label{fig:d0.9}
\end{minipage}
\end{figure}

\begin{figure}[h]
\epsfig{figure= 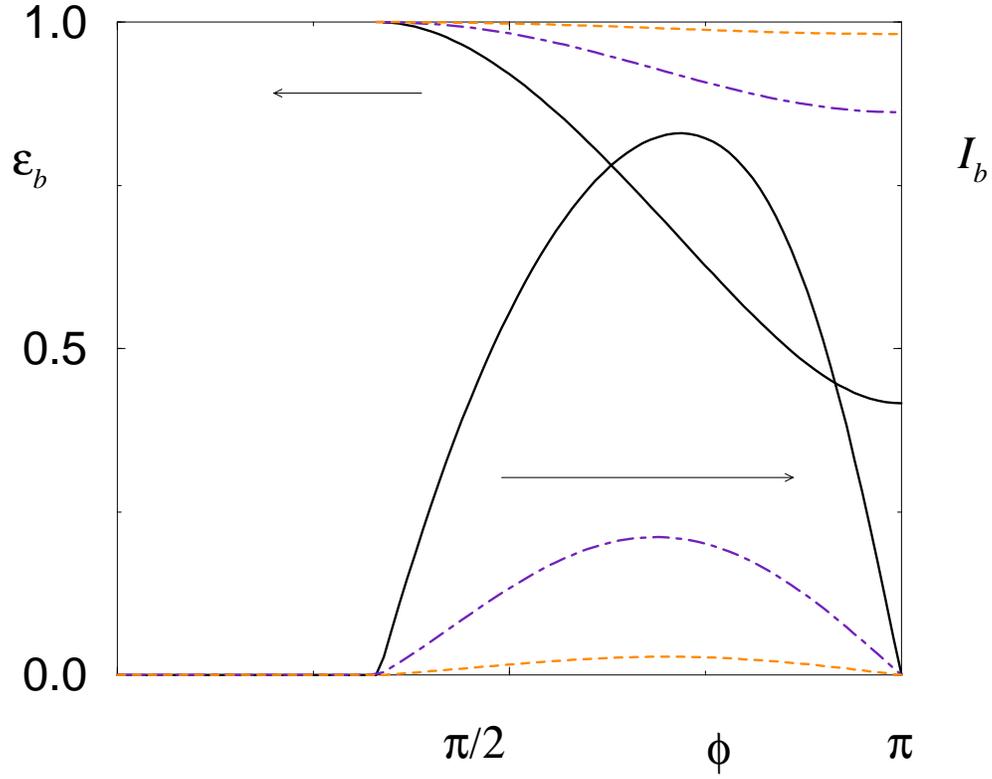,width=5in,angle=-90}
\vskip 0.4 cm
\begin{minipage}{\textwidth}
\caption[]{ Bound state energy $\epsilon_b$ (in unit
of $|\Delta_l|$, left axis), and 
the bound state contribution to the current
(in unit of $e |\Delta_l| / \hbar $, right axis) 
as a function of $\phi$. 
$|\Delta_r|/|\Delta_l| = 2.0$,
while $D$'s are given by
$0.9$ (full line), $0.5$ (dot-dashed),
and $0.2$ (long-dashed). }

\label{fig:z0.5}
\end{minipage}
\end{figure}

\begin{figure}[h]
\epsfig{figure= 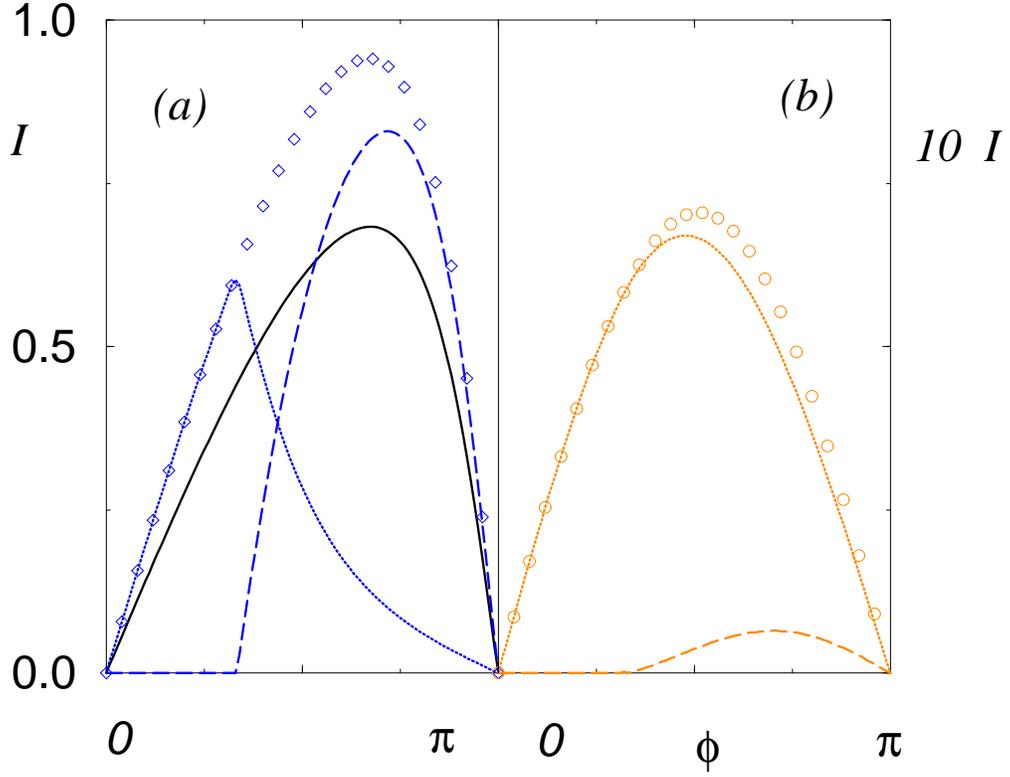,width=5in,angle=-90}
\vskip 0.4 cm
\begin{minipage}{\textwidth}
\caption[]{ Current-phase dependence
for (a): $D=0.9$, $|\Delta_r|/|\Delta_l| = 2.0$.
$T = 0$, and
currents are in units of $ e |\Delta_l| \over \hbar$.
Long-dashed:  $I_b$, dotted: continuum contribution,
symbols:  total.   The current
for $|\Delta_r| = |\Delta_l|$ is also shown
(full line) for comparison.
(b): Same as (a) except $D=0.1$.
  Note the difference in scale for $I$. }
\label{fig:cura}
\end{minipage}
\end{figure}

\begin{figure}[h]
\epsfig{figure= 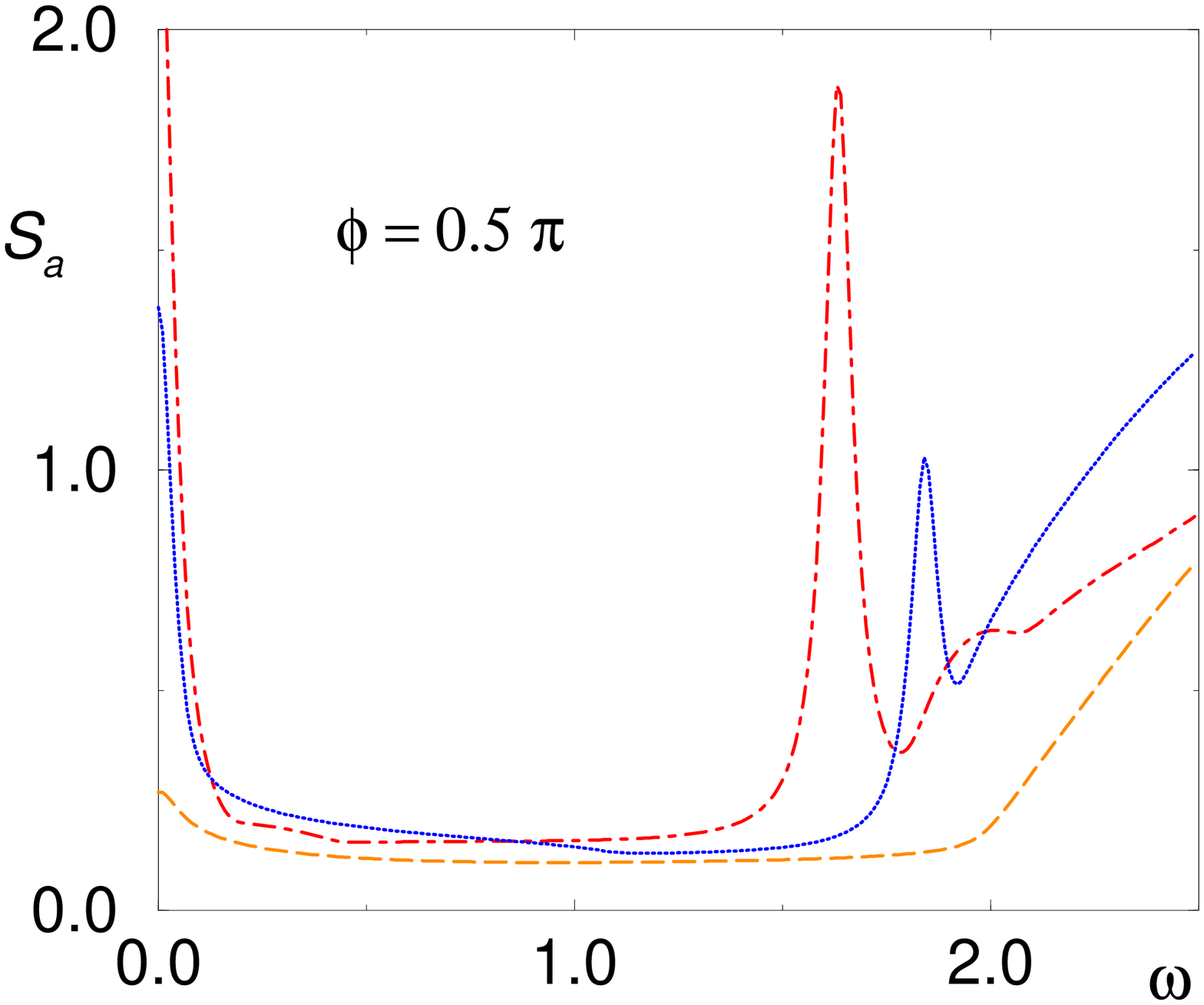,width=5in,angle=-90}
\vskip 0.4 cm
\begin{minipage}{\textwidth}
\caption[]{ $S_a(\omega)$
for $D=0.9$, $\phi= 0.5 \pi$, $T / |\Delta_l|= 0.3$, and
damping coefficient $\eta = 0.02 |\Delta_l|$.
$|\Delta_r|/|\Delta_l|$ are given by,
dot-dashed: $1.0$, dotted: $2.0$, long dashed $10.0$.
$S_a$ in unit of $ e^2 |\Delta_l| \over \hbar$,
$\omega$ in unit of $|\Delta_l|$. }

\label{fig:noise}
\end{minipage}
\end{figure}



\begin{thebibliography}{10}
\bibitem{notes}  We are following the suggestions of
\cite{Gavish01} to study an asymmetrically defined current-current
correlation function which directly relates to a certain
(here absorption) processes.  The spectral density
$S (\omega)$ defined in, e.g., \cite{AverinL96} with
$\hat I (t) \hat I (0)$ replaced by 
${1 \over 2} ( \hat I (t) \hat I (0) + \hat I (0) \hat I (t) )$
is given by $S (\omega) = { 1 \over 2} ( 1 + e^{- \omega / T} ) S_a (\omega)$.
\bibitem{Appendix}
Here, for completeness, we provide the definitions
for the envelope functions $\hat C$'s and explain how
they can be obtained.  Near the interface the Nambu
Green's function $\hat G$ can be expressed in the form
\cite{MRS}
\be
\hat \tau_3 \hat G (x,t,x',t')
  = { 1 \over p/m} \sum_{j,j'} \hat C_{j j'} (x,t,x',t')
     e^{ i p ( j x - j' x') }
\ee
such that $\hat C_{j j'}$ are slowly varying functions of
$x$ and $x'$. Here $j$, $j'$ are $\pm $.  
The $\hat C_{j j'} (\epsilon)$ are the Fourier transform ( in time)
of $\hat C_{j j'} (t)$ which are in turn given by
$ { {\rm lim} \atop { x' \to x + 0_{+}} } \hat C_{j j'} 
   ( x, t, x', 0)$.

$\hat C_{++}$ and $\hat C_{--}$ were already given in text.
From eq (54) of \cite{MRS} ( with $l$ and $r$ interchanged),
we have 
\be
r \hat C_{l,+-} + r^{*} \hat C_{l,-+} =
\hat C_{l,++} + |r|^2 \hat C_{l,--} - D \hat C_{r,++}
\label{bc}
\ee
We can eliminate $\hat C_{l,+-}$ by multiplying this
equation by $( \hat C_{l,++} - i )$, since \cite{MRS}
$ (\hat C_{l,++} - i ) \hat C_{l,+-} =  0$.
We thus obtain an expression for
$ (\hat C_{l,++} - i ) \hat C_{l,-+}$.  
On the other hand 
$ (\hat C_{l,--} + i ) \hat C_{l,-+} =  0$.
Subtracting these two equations, we obtain an 
equation from which we can solve for $\hat C_{l,-+}$,
since $ ( \hat C_{l,++} - \hat C_{l,--} - 2 i)$
is invertible ({\it c.f.} \cite{Yip97}).
$\hat C_{l,+-}$ can be obtained in a similar manner.
\end{thebibliography}
\end{document}